\documentclass[preprint,12pt]{elsarticle}
\usepackage{amssymb}
\usepackage{epsfig}
\usepackage{amsmath, amsthm}
\usepackage{graphicx}
\journal{?}

\begin{document}

\begin{frontmatter}

\title{Effects of Membrane Morphology on the Efficiency of Direct Contact Membrane Distillation.}

\author{Gavin A. Buxton}

\address{Science Department, Robert Morris University, Moon Township, PA 15108, US. \\Email: buxton$@$rmu.edu}

\begin{abstract}
A computer simulation is used to predict the effects of membrane morphology on the thermal efficiency of direct contact membrane distillation. 
The mass transfer through the porous microstructure and the heat conduction through the membrane are both related by the membrane morphology.
The interrelated tortuosity of the porous structure and solid phase influences the mass transfer and thermal conductivity, respectively.
The effects of varying the morphology are elucidated and introducing a lattice structure, which tailors the morphology, can significantly increase thermal efficiency.
A three-layer system is also simulated, where the pore size in the middle layer can be increased without significantly increasing the risk of membrane pore wetting.
Three-layer systems that possess a lattice morphology are found to result in thermal efficiencies around 20\% higher than random morphologies.
\end{abstract}


\end{frontmatter}

\section{Introduction}
\label{introduction}

Population growth and expanding urbanization is stressing fresh water supplies, causing disease, wars, famine and population displacement \cite{eliasson2015, mekonnen2016}.
Furthermore, increasing industrialization and agricultural practices are polluting our existing water sources, while global warming is exacerbating water shortages \cite{wu1999, vorosmarty2000, daliakopoulos2016}.
While water conservation, reuse and reclamation can mitigate these problems, to alleviate these water scarcity issues there is a growing need to inexpensively desalinate seawater \cite{service2006}. 
While seawater desalination typically uses reverse osmosis or conventional thermal distillation, membrane distillation could play a part in meeting our future freshwater needs \cite{pangarkar2016}. 

Membrane distillation is a thermal desalination technology where the feed water is warmed and flows past a hydrophobic membrane that only allows the vapor molecules to pass through to the cool distillate water on the other side.
Membrane distillation has been known for over 50 years \cite{bodell1963, findley1967}, but has experienced limited commercial success because it is generally not as economically viable, or as energetically favorable, as alternative desalination technologies \cite{pangarkar2016}.
That said, membrane distillation is attracting attention for small scale and niche applications.
In particular, membrane distillation can produce high purity distillate, with 100\% rejection of nonvolatile compounds, and could be competitive for certain applications \cite{banat1998, alkhudhiri2012}.

The membranes in membrane distillation have larger pores, than the membranes used in reverse osmosis desalination \cite{alkhudhiri2012}, and the fluid is not forced through the pores in its liquid state \cite{duong2015}, which is believed to result in less fouling; a major drawback of the reverse osmosis process \cite{termpiyakul2005, goosen2005}. 
Microbial growth and scaling, while not as pronounced as in reverse osmosis membranes, can cause a decline in permeate flux and a reduction in liquid entry pressure, causing membrane pore wetting and feed water to flow through the membrane \cite{gryta2002, alklaibi2005, warsinger2015, duong2015, duong2017b}.
However, even for extremely high salt concentrations scaling can be easily controlled by regular membrane flushing \cite{nghiem2011}.

The feed temperature is also relatively low, and the pressure throughout the system is much lower than pressure-driven membrane processes (like reverse osmosis), resulting in less corrosive, less mechanically-demanding, less expensive equipment, and fewer safety concerns \cite{alkhudhiri2012}.
The energetic costs of heating the feed in membrane distillation could be provided by low-grade thermal energy sources.
For example, the feed temperatures typically range from 40\textsuperscript{o}C to 80\textsuperscript{o}C, which can correspond with temperatures from thermal solar collectors or geothermal energy sources \cite{duong2015, duong2017}.
Coupling the membrane distillation process with renewable energy sources, or existing sources of waste heat, could further reduce the environmental impacts of the desalination process \cite{adnan2012, tarnacki2012}.
The portability and reliability of membrane distillation could also meet the need to purify the significant volumes of wastewater produced locally by hydraulic fracturing; the flaring of natural gas could provide waste heat for this process \cite{glazer2014}.
There might be many situations were the high energy input needed for heating the feed water could be obtained at little to no cost \cite{jansen2013}.

In membrane distillation the feed water is heated and flows on one side of the membrane.
In the case of direct-contact membrane distillation, colder distillate water flows directly on the other side of the membrane; although alternative configurations of membrane distillation capture and condense the pure water vapor differently.
The hydrophobic microporous membrane has a high liquid entry pressure, meaning that water cannot penetrate the membrane pores, and vapor-liquid interfaces exist at the membrane surface (at the opening of the pores).
Furthermore, because of the difference in temperature between the feed and distillate water, a vapor pressure difference occurs across the membrane.
In other words, the water evaporates at the warm interface, diffuses through the membrane, and condenses at the cold interface resulting in a flux of (purified) water across the membrane.
The process is generally much more efficient when the feed temperature is higher \cite{alklaibi2005, duong2017}, but is also dependent on the feed flow rate (and distillate flow rate, to a lesser extent)\cite{ohta1990, alklaibi2005}, the salinity of the feed (to a small extent)\cite{alklaibi2005}, and membrane properties.

Besides direct-contact membrane distillation, the more common membrane distillation configurations include air gap, vacuum and sweeping gas membrane distillation \cite{feng2008, mengual2004, khayet2003}.
Air gap membrane distillation incorporates a layer of stagnant air between the membrane and the condensation surface, which limits heat and mass transfer; limiting thermal conduction is advantageous, but limiting the flow of water vapor is a disadvantage \cite{feng2008}.
In vacuum membrane distillation a vacuum is created in the distillate side of the membrane and condensation occurs external to the membrane system; this eliminates conductive heat losses across the membrane \cite{mengual2004, alkhudhiri2012}.
Inert gas is used to transport the water vapor to an external condenser in the sweeping gas membrane distillation, which enhances the transfer of water vapor \cite{khayet2003}.
However, while direct contact membrane distillation may incur greater heat losses than other configurations, it has the simplest arrangement.

As heat and mass is transfered across the membrane, the temperature and salt concentration in layers adjacent to the feed membrane surface can increase \cite{phattaranawik2003, termpiyakul2005, alkhudhiri2012, duong2017b}.
The emergence of a temperature polarization (a difference in temperature between the membrane and bulk fluid) is thought to be more detrimental than a concentration polarization (a difference in salinity between the membrane and bulk fluid) \cite{martinez1999}.
The inclusion of various obstacles \cite{phattaranawik2001, phattaranawik2003c, martinez1998, yun2011, yang2012, yu2012}, surface roughness \cite{ho2013}, and gas bubbles \cite{chen2013} in the feed flow have been used to promote turbulence in the feed flow and decrease temperature (and presumably concentration) polarization.
Decreasing temperature polarization increases the difference in water vapor pressure, and hence increases the water flux.

The microporous hydrophobic polymeric membranes can be fabricated using a variety of processes, and a low resistance to mass transfer and a low thermal conductivity are desirable.
The heat and mass transfer through these membranes is a complex process depending on the membrane thickness, porosity, pore size and tortuosity \cite{khayet2004, khayet2011}.
Mass transfer is increased with decreasing thickness, but heat loss is also increased with decreasing thickness \cite{khayet2011, alkhudhiri2012, adnan2012}. 
An optimum thickness, therefore, maximizes mass transport while minimizing heat transfer \cite{lagana2000, khayet2011}.
A high porosity is found to result in higher fluxes and lower thermal conductivities, but membranes that are too porous might lack mechanical stability \cite{khayet2011, alkhudhiri2012, adnan2012}.
The size of the pores can vary from around 100 nm up to 1 $\mu$m \cite{alkhudhiri2012}, with the pore size dictating not only the nature of mass flow but also the liquid entry pressure (required to prevent membrane wetting) \cite{khayet2011}.
The tortuosity is the path length through the membrane relative to thickness.
A small tortuosity is thought to be favorable, as the straight pores through the system can facilitate mass transfer \cite{khayet2011}.
However, the optimum morphology for mass transfer through these membranes can be complicated by the need to limit thermal conductivity through the same structure.
While a low tortuosity is considered desirable in membrane distillation, as it will facilitate mass transfer through the membrane, it might also be expected to encourage heat transfer, as this would usually correspond to solid (more thermally conducting) material forming a more direct path through the membrane.
The mass transfer through the porous structure and the heat transfer through the solid matrix are therefore coupled through the same morphology, and there have been few attempts to couple such cross-property connections \cite{sevostianov2010, khayet2010}.
The ideal membrane structure for membrane distillation, therefore, is not always obvious.
In the current paper the effects on thermal efficiency of varying membrane attributes that are coupled to membrane morphology are considered.
Furthermore, the benefits of three-layer membranes \cite{wang2011, wang2015, amin2017} are also investigated, where the pore size in a middle layer can be increased without significantly increasing the membranes susceptibility to pore wetting.
We also consider a lattice network geometry and show that three-layer lattice network morphologies can significantly improve thermal efficiencies.

\section{Methodology}
\label{methodology}

Gas transport through the porous membrane depends on the size and tortuosity of the pores.
The Knudsen number, $K_n$, is defined as the ratio of the mean free path, $\lambda$, of the gas molecules to the pore diameter, $d_p$ \cite{hitsov2015}.
Knudsen-diffusion occurs at large values of $K_n$ (greater than one), and the gas diffuses through the porous membrane more through wall-molecule collisions than molecule-molecule collisions.
When $K_n < 0.01$ the mass transfer occurs through molecular-diffusion, and the gas diffuses primarily via molecule-molecule collisions.
For the smaller values of $K_n$ the fluid dynamics can be increasingly described by Navier-Stokes flow, as no-slip boundary conditions become appropriate to describe the boundary flow \cite{zhang2012}. 
For intermediate values of $K_n$, however, both Knudsen- and molecular-diffusion can occur simultaneously, which is considered to be the case in membrane distillation \cite{khayet2004}.

For Knudsen-diffusion through porous media the mass transfer coefficient is of the form \cite{phattaranawik2003, khayet2004, qtaishat2008}
\begin{equation}
C_m = \frac{2 \epsilon r}{3 \tau \delta} \left(\frac{8M}{\pi RT}\right)^{\frac{1}{2}}
\end{equation}
where $\epsilon$ is the porosity, $r$ is the average pore radius, $\tau$ is the tortuosity, $\delta$ is the membrane thickness, $M$ is the molecular weight of water, $R$ is the gas constant, and $T$ is the temperature.

The molecular-diffusion is described by
\begin{equation}
C_m = \frac{\epsilon PD M}{\tau \delta P_a RT}
\end{equation}
where $P$ is the pressure inside the pore, $D$ is the diffusion coefficient, and $P_a$ is the air partial pressure.
This describes the mass transport when the dominant interactions inside the pores are molecule-molecule collisions \cite{khayet2004, qtaishat2008, hitsov2015}.

In the transition region, where both Knudsen- and molecular-diffusion occurs, the mass transfer coefficient is given by
\begin{equation}
C_m = \left[ \frac{3 \tau \delta}{2 \epsilon r} \left(\frac{\pi RT}{8M}\right)^{\frac{1}{2}} + \frac{\tau \delta P_a RT}{\epsilon PD M}\right]^{-1}
\end{equation}
which combines Knudsen- and molecular-diffusion mass transfer coefficients \cite{khayet2004}.
This has been found to best describe the gas transport in membrane distillation  \cite{khayet2004, termpiyakul2005, andrjesdottir2013}.

In the above equations the pore radius is taken to be a constant, and the effects of pore-size distribution are neglected \cite{hitsov2015}. 
In real pores, the different mechanisms of gas transport may occur to different extents in different pores \cite{khayet2001, khayet2004}.
Khayet \emph{et al.} and Phattaranawik \emph{et al.} have investigated the mass transport in membranes with pore-size distributions \cite{phattaranawik2003b, khayet2004} and the effects of pore-size distribution was found to be insignificant \cite{phattaranawik2003b}, although this may only be true for the low variations in pore-sizes found in commercial membranes \cite{khayet2004}.

The thermal conductivity of the membrane dictates heat losses as heat conducts from the warm feed to the cold distillate. 
While the presence of water in wetted pores \cite{gryta2012} and thin air films adjacent to the membrane \cite{dumee2013, hitsov2015} might add additional complexity, the thermal conductivity of the membrane is usually estimated from the thermal conductivities of the polymer and gas phases, and the membrane morphology.
The thermal conductivity of the gas, $k_g$, is treated as a single component (rather than separating air and water vapor) as the thermal conductivities of both air and water vapor are an order of magnitude smaller than the polymer phase \cite{phattaranawik2003, alkhudhiri2012}.
The temperature dependence of the polymer thermal conductivity, $k_s$, will depend on the material, but generally increases with temperature \cite{hitsov2015}. 

The simplest bounds to the composite thermal conductivity are the Voigt-Reuss bounds (see Fig 1a).
The Voigt upper bound is the limiting case when the two phase are aligned parallel to the temperature gradient; as would be the case, for example, if the pores consisted of cylinders directly through the membrane.
The membrane thermal conductivity is estimated as
\begin{equation}
k_m = \epsilon k_g + (1-\epsilon) k_s
\end{equation}
where $\epsilon$ is porosity.
This form has been used to describe the thermal conductivity in membrane distillation in a number of studies \cite{martinez1998, martinez1999, adnan2012, alkhudhiri2012, gryta2012, andrjesdottir2013}, but is only appropriate for systems where the pore channels are approximately straight and perpendicular to the membrane surface \cite{zhang2012}.
The Reuss lower bound has also been used to describe the thermal conductivity \cite{phattaranawik2003, qtaishat2008, alkhudhiri2012} which uses the volume-average of the reciprocal of the conductivities 
\begin{equation}
k_m = \frac{k_g k_s}{k_s \epsilon + k_g (1 - \epsilon)}
\end{equation}
This is the limit when the pores are aligned parallel to the membrane surface \cite{hitsov2015}.

The Voigt-Reuss-Hill approximation is the average of the upper and lower Voigt-Reuss bounds, as shown in Fig. 1a.
This is a specific example of a hybrid Voigt-Reuss mixture, for which the thermal conductivity is given as 
\begin{equation}
k_m = \alpha \left[ \epsilon k_g + (1-\epsilon) k_s\right] + (1 - \alpha) \left[ \frac{k_g k_s}{k_s \epsilon + k_g (1 - \epsilon)}\right]
\end{equation}
where $\alpha$ is the hybridizing parameter (a number between zero and one) and is taken to be $\frac{1}{2}$ in the Voigt-Reuss-Hill approximation.

Other, narrower, bounds on the thermal conductivity exist. 
For example, the Hashin-Shtrikman bounds are often used when only one phase is continuous \cite{hashin1963}.
The Maxwell Type I equation has been used to capture the thermal conductivity in membrane distillation \cite{garcia2004, hitsov2015}.
The form of this is 
\begin{equation}
k_m = \frac{k_g (3 k_s - 2 \epsilon (k_s-k_g))}{3 k_g + \epsilon(k_s - k_g) }
\end{equation}
which gives an approximation in between the Voigt-Reuss bounds, as shown in Fig 1a.
However, it is worth noting that the applicability of these different approximations will depend on the porous structure of the membrane, and the form could be chosen to best fit experimental results.

A theoretical model is used to predict the effects of different membrane properties in direct contact membrane distillation \cite{khayet2004, duong2017}.
The mass transfer is described by the water flux
\begin{equation}
J = C_m (P_{mf} - P_{md})
\end{equation}
where $C_m$ is the mass transfer coefficient, and $P_{mf}$ and $P_{md}$ are the water vapor pressures at the feed and distillate membrane surfaces, respectively.
The water vapor pressure decreases with salinity 
\begin{equation}
P = x_{water} \left( 1 - 0.5 x_{salt} - 10 x_{salt}^2\right) P_0
\end{equation}
where $x_{water}$ and $x_{salt}$ are the molar fraction of water and salt, respectively. 
The molar fraction of salt depends on the salinity, $S$, and is given by
\begin{equation}
x_{salt} = \frac{S M_w}{M_s(1-S) + S M_w}
\end{equation}
where $M_w$ and $M_s$ are the molecular weights of water and salt, respectively.
The vapor pressure of pure water, $P_0$, has the following temperature dependence
\begin{equation}
P_0 = \exp\left(23.1964-\frac{3816.44}{T -46.13}\right)
\end{equation}
where $T$ is temperature and the nonlinearity of Antoine's equation above predicts that the performance of membrane distillation will depend more on variations in the feed temperature than variations in the distillate temperature \cite{alkhudhiri2012}.

The mass transfer coefficient, assuming combined Knudsen- and molecular-diffusion mechanisms of gas transport, is of the form
\begin{equation}
C_m = \left[ \frac{3 \tau \delta}{2 \epsilon r} \left(\frac{\pi RT}{8M}\right)^{\frac{1}{2}} + \frac{\tau \delta P_a RT}{\epsilon PD M}\right]^{-1}
\end{equation}
where $\epsilon$ is the porosity, $r$ is the average pore radius, $\tau$ is the tortuosity, $\delta$ is the membrane thickness, $M$ is the molecular weight of water, $R$ is the gas constant, $T$ is the temperature, and $P_a$ is the air partial pressure.
The product of pressure inside the pore, $P$, and the diffusion coefficient, $D$ is given by \cite{duong2017}
\begin{equation}
PD = 1.895 \times 10^{-5}\,T^{2.072}
\end{equation}
Assuming that there is no heat loss to the surroundings, and that the heat flux through the system (from bulk feed though to bulk distillate) is constant, the following equations for the temperatures at the membrane surfaces can be obtained \cite{khayet2004}
\begin{equation}
T_{mf} =\frac{T_{bf} h_f + h_m \left( T_{bd} + T_{bf} \frac{h_f}{h_d}\right) - J \Delta H_{\nu}}{h_f \left(1 + \frac{h_m}{h_d}\right) + h_m}
\end{equation}
and
\begin{equation}
T_{md} =\frac{T_{bd} h_d + h_m \left( T_{bf} + T_{bd} \frac{h_d}{h_f}\right) + J \Delta H_{\nu}}{h_d \left(1 + \frac{h_m}{h_f}\right) + h_m}
\end{equation}
where $T_{mf}$ and $T_{md}$ are the temperatures at the feed and distillate membrane surfaces, respectively. $T_{bf}$ and $T_{bd}$ are the temperatures in the bulk feed and distillate streams, respectively. 
The heat transfer coefficients between the bulk and membrane layers for the distillate and feed streams are given by
\begin{equation}
h_d = \frac{Nu_d k_d}{d_h}
\end{equation}
and
\begin{equation}
h_f = \frac{Nu_f k_f}{d_h}
\end{equation}
The conductive heat transfer coefficient across the membrane is the thermal conductivity of the membrane divided by the membrane thickness.
\begin{equation}
h_m = \frac{k_m}{\delta}
\end{equation}
For a flat sheet module at laminar flow the Nusselt number, $Nu$, is of the form \cite{hitsov2015, duong2017}
\begin{equation}
Nu = 0.097 Re^{0.73} Pr_b^{0.13} \left(\frac{Pr_b}{Pr_m}\right)^{0.25}
\end{equation}
where the Reynolds number, $Re$, and Prandtl number, $Pr$, can be obtained from
\begin{equation}
Re = \frac{\rho u d_h}{\mu}
\end{equation}
and
\begin{equation}
Pr = \frac{\mu C_p}{k}
\end{equation}
where $\rho$, $\mu$, $k$ and $C_p$ are the density, dynamic viscosity, thermal conductivity, and specific heat capacity of the feed and distillate streams.
$u$ is the stream velocity and the hydraulic diameter, $d_h$, is given as
\begin{equation}
d_h = \frac{4 W H}{2 W + 2H}
\end{equation}
where $W$ and $H$ are the width and height of the fluid channels.

The simulation domain is discretized in the flow direction \cite{duong2017}.
The mass transfer rate at the i\textsuperscript{th} location along the channel (node) is
\begin{equation}
dm_i = J_i\, W\, dx
\end{equation}
where $J_i$ is the water flux at the i\textsuperscript{th} node, and $dx$ is the spatial step (1 mm in the current model).
Similarly, the heat flux through an area $W\, dx$ at the i\textsuperscript{th} node is
\begin{equation}
dQ_i = \left(J_i \Delta H_{\nu} + \frac{k_m}{\delta} \left( T_{mf} - T_{md}\right)\right)\, W\, dx
\end{equation}
where $k_m$ is the thermal conductivity of the membrane, $\delta$ is the membrane thickness, and the latent heat of vaporization, $\Delta H_{\nu}$, is given by \cite{khayet2010}
\begin{equation}
\Delta H_{\nu} = 1.7535 \, \frac{T_{mf} + T_{md}}{2} + 2024.3
\end{equation}

As the fluid flows down the channel, and the mass of water diffuses from the feed to the distillate, the salinity of the feed is going to increase as the mass of the feed fluid decreases.
The decrease in feed mass flow rate is
\begin{equation}
m_{f, i+1} = m_{f,i} - dm_i
\end{equation}
where $m_{f,i}$ and $m_{f,i+1}$ are the feed mass flow rates of water at the i\textsuperscript{th} and i+1\textsuperscript{th} nodes, respectively. 
The salinity of the i+1\textsuperscript{th} node is 
\begin{equation}
s_{f,i+1} = \frac{m_{f,i} s_{f,i}}{m_{f,i+1}}
\end{equation}
where $s_{f,i}$ is the salinity at the previous node.

The temperature of the bulk feed is reduced as heat is transferred across the membrane.
\begin{equation}
T_{bf,i+1} = \frac{m_{f,i} Cp_{f,i} T_{bf,i} - dQ_i}{m_{f,i+1} Cp_{f,i}}
\end{equation}
where $T_{bf,i}$ and $T_{bf,i+1}$ are the temperatures in the bulk feed stream at the i\textsuperscript{th} and i+1\textsuperscript{th} nodes, respectively. 
$Cp_{f,i}$ is the specific heat of the fluid at the i\textsuperscript{th} node \cite{duong2017}. 

The distillate will acquire mass in the direction it is flowing, which could be either in the same or in the opposite direction of the feed stream \cite{duong2017}. 
For co-current flow (in the same direction as the feed) the mass flow rate and temperature equations are given by
\begin{equation}
m_{d, i+1} = m_{d,i} + dm_i
\end{equation}
and
\begin{equation}
T_{bd,i+1} = \frac{m_{d,i} Cp_{d,i} T_{bd,i} + dQ_i}{m_{d,i+1} Cp_{d,i}}
\end{equation}
where variables have the same meaning as in the feed equation, but with the subscript \emph{d} representing the distillate stream.
For counter-current flow (in the opposite direction as the feed) the mass and temperature equations are given by
\begin{equation}
m_{d,i+1} = m_{d,i} - dm_i
\end{equation}
and
\begin{equation}
T_{bd,i+1} = \frac{m_{d,i} Cp_{d,i} T_{bd,i} - dQ_i}{m_{d,i+1} Cp_{d,i}}
\end{equation}
Here we consider counter current flow as this has generally been found to be more efficient.
In this manner the mass flow rate and temperature of the fluid streams can be captured along the length of the system.

In the above equations the heat and mass transfer are coupled through the vaporization and transport of warm water molecules across the membrane.
However, they are also coupled by the morphology of the porous structure.
In particular, a correlation between the mass transfer coefficient and thermal conductivity might be expected.
Recall that the thermal conductivity can be described using the hybrid Voigt-Reuss equation
\begin{equation}
k_m = \alpha \left[\epsilon k_g + (1 - \epsilon) k_s\right] + (1-\alpha) \left[\frac{k_g k_s}{\epsilon k_s + (1 - \epsilon)k_g} \right]
\end{equation}
While the tortuosity is often described as the path length through a tortuous passage divided by thickness, an alternative definition is of the form \cite{perkins1956, sevostianov2010, garcia2012}
\begin{equation}
\frac{k_s}{k_m} = \frac{\tau_s^2}{1 - \epsilon}
\end{equation}
which relates the tortuosity of the solid phase, $\tau_s$, to the thermal conductivity of the solid phase.
Comparing this to the hybrid Voigt-Reuss equation (for just the solid phase) gives
\begin{equation}
\tau_s = 1/\sqrt{\alpha}
\end{equation}
where $\alpha$ is the hybridizing parameter.
For a bicontinuous structure, Sevostianov and Shrestha \cite{sevostianov2010} obtained the following approximation for the cross-correlation between the tortuosity of the different phases
\begin{equation}
\frac{1}{\tau} = 1 - \frac{1-\epsilon}{\epsilon} \left(1-\frac{1}{\tau_s}\right)
\end{equation}
where $\tau$ is the tortuosity of the pores. 
This approximation assumes a symmetric distribution of pore sizes, and at $\epsilon = 0.5$ it is assumed $\tau = \tau_s$.
This explicit relationship has been verified through simulations of fluid permeability and electrical conductivity in porous structures \cite{garcia2012}.
We can, therefore, obtain the following expression relating the tortuosity in the membrane to the hybridizing parameter
\begin{equation}
\tau = \frac{\epsilon}{\epsilon - (1-\epsilon)\left(1-\sqrt{\alpha}\right)}
\end{equation}
The relationship for different porosities is shown in Fig. 1b.
Regardless of porosity, the tortuosity of the structure is predicted to go to one (straight channels through the membrane) as the hybridizing parameter goes to one (Voigt upper bound).
This is expected as the Voigt upper bound corresponds with perfectly aligned structures perpendicular to the membrane.
As the hybridizing parameter is reduced the tortuosity increases, more for systems with lower porosity.
For a porosity of 0.5, the tortuosity will go to infinity as the hybridizing parameter goes to zero, and the pores are aligned parallel to the membrane.
For higher porosities, however, the tortuosity does not go to infinity as the hybridizing parameter goes to zero; in other words, as the tortuosity of the solid goes to infinity it is predicted that the tortuosity of the gas phase does not.
The porous structure is assumed to always have a percolating pathway through the membrane in systems with higher porosities.
However, this explicit relationship is only an estimate based on randomly distributed interconnected phases, and does not take into consideration more tailored structures.

To characterize the performance of direct contact membrane distillation systems, the temperature polarization and thermal efficiency are often used.
The temperature polarization is defined as the difference in temperature across the membrane relative to the difference in bulk temperatures.
\begin{equation}
\Psi = \frac{T_{mf} - T_{md}}{T_{bf} - T_{bd}}
\end{equation}
The thermal efficiency is defined as
\begin{equation}
\Pi = \frac{J_i \Delta H_{\nu}}{J_i \Delta H_{\nu} + \frac{k_m}{\delta} \left( T_{mf} - T_{md}\right)}
\end{equation}
The thermal efficiency increases with increasing water flux, and increases in systems with lower conductive heat losses through the membrane. 
It is this interdependence which makes these systems complex; the microstructural properties that might increase water flux (increasing thermal efficiency) might also be expected to increase thermal conduction (decreasing thermal efficiency).
The questions remain: is it more advantageous to promote mass transfer or limit thermal conductivities?
Are there morphologies which might achieve both?

\section{Results and Discussion}
\label{results}

\subsection{The Effects of Membrane Attributes}
\label{results1}

The feed and distillate mass flows are initially 1000 kg/h (0.35 m/s), the feed input salinity is 0.035 (or 35 g/L) and the distillate input temperature is 20\textsuperscript{o}C. 
The thickness of the membrane is 100 $\mu$m and the pore radius is 0.2 $\mu$m.
The length of the system is 1.5 m, the width is 0.4 m and the height of the channels are 2 mm.
The effects of varying these parameters are not considered in the current study, but increasing the length of the system is expected to decrease efficiency as the temperature difference across the membrane will decrease with length.

Figure 2 shows the thermal efficiency as a function of the hybridizing parameter in the Voigt-Reuss mixture, $\alpha$, and the tortuosity of the membrane pores, $\tau$.
The effects of changing the porosity from 0.5 to 0.8 are depicted.
The input feed temperature is 60\textsuperscript{o}C.
The scale of the thermal efficiencies observed increase from a range of 0.25 to 0.65 in the system with a porosity of 0.5 to a range of 0.45 to 0.8 as the porosity of the membrane increases to 0.8.
This is expected, as increasing porosity is expected to increase water flux while decreasing thermal conduction.
It is worth noting that the model assumes no other heat losses in the system.
The higher efficiencies are predicted to occur at lower values of $\alpha$ and lower values of $\tau$. 
Lower values of $\alpha$ correspond with the Reuss lower bound and lower thermal conductivities, while the lower values of $\tau$ increase the mass transfer coefficient and increase water flux.
However, it may be unreasonable to assume that the microstructure can simultaneously accommodate these extremes.
The explicit relation derived from Sevostianov and Shrestha \cite{sevostianov2010}, between $\tau$ and $\alpha$, is shown as a black line superimposed over the contour plots.
In Fig 2a while this line avoids the area of highest efficiency, it also avoids the area of lowest efficiency, and it appears that the efficiency might not be too sensitive to the microstructure at a porosity of 0.5.
As the porosity increases, the line passes through higher efficiencies at lower values of $\alpha$ and higher values of $\tau$.
In other words, in these bicontinuous structures it is predicted that the more tortuous morphologies are more efficient, and that limiting heat transfer may be favorable over increasing mass transfer.

Figure 3 shows the thermal efficiency, as a function of the hybridizing parameter and the tortuosity, in systems with feed temperatures of 40\textsuperscript{o}C, and 80\textsuperscript{o}C. 
In both systems the porosity is fixed at 0.65 and the other parameters remain the same.
As the temperature increases the overall efficiency in the systems increases, as expected as the water vapor pressure at the feed membrane increases which drives water flux across the membrane. 
The profiles, however, look similar and it would appear that the optimized microstructure might not depend on the feed temperature.
Fig. 3c depicts the thermal efficiency in systems with feed temperatures of 40\textsuperscript{o}C, 60\textsuperscript{o}C, and 80\textsuperscript{o}C. 
The thermal efficiency is plotted as a function of the hybridizing parameter, $\alpha$.
$\tau$ is assumed to depend on $\alpha$ following the explicit relation derived from Sevostianov and Shrestha \cite{sevostianov2010}. 
In other words, the thermal efficiencies are calculated for values of $\tau$ and $\alpha$ that correspond with the solid lines in Fig. 3a and 3b. 
While increasing the feed temperature can significantly increase the thermal efficiency, as expected, the effect of morphology also plays a significant role for membranes consisting of randomly distributed interconnected phases.
It is predicted that varying the morphology can change the efficiency by several percent, and that the peak efficiency occurs close to the Reuss lower bound.

The hybridizing parameter and tortuosity can be varied and the maximum thermal efficiency can be found in systems with different porosities. 
In other words, the system can be optimized as a function of $\alpha$ and $\tau$ (which is now taken to be a function of $\alpha$).
For the randomly distributed interconnected structures the optimized morphology is predicted to depend on the membrane porosity.
Figure 4 shows the optimized hybridizing parameter, tortuosity and thermal efficiency as a function of porosity. 
As porosity increases, the optimized hybridizing parameter decreases, and the optimized structure moves from one closer to the Voigt bound (with channels aligned perpendicular to the membrane) to the Reuss bound (with the pores aligned more along the membrane).
The tortuosity depends on the porosity, with lower tortuosities expected to occur in systems with higher porosities, and the optimized tortuosity shows a maximum at a porosity of 0.62. 

In Fig 4b the predicted thermal efficiency increases with porosity, as expected, as increasing porosity increases both mass transfer and the resistance to heat conduction.
The optimized thermal efficiency is compared to the thermal efficiency in systems obeying either the Voigt bounds ($\alpha = 1$) or the Reuss bounds ($\alpha = 0$).
In the Reuss system the thermal efficiency goes to zero as porosity goes to 0.5. 
The Voigt bounds are closer to the optimized system at these lower porosities, but at higher porosities the optimized system approaches the Reuss bounds (which severely reduces thermal conduction). 
At these higher porosities, however, there is less variation in thermal efficiency in systems with different morphologies. 

\subsection{Three-Layer Membranes}
\label{results2}

The incorporation of a layer of polymer in the middle of the membrane with varying thickness and increased pore radius is considered.
The overall membrane thickness is maintained at 100 $\mu$m and the temperature of the feed is 60\textsuperscript{o}C.
The membrane layers on either side (next to the feed and distillate) are kept the same, with a pore radius of 0.2 $\mu$m, to ensure a sufficient liquid entry pressure to inhibit pore wetting.
The tortuosity as a function of pore radius and thickness of the central layer are depicted in Figure 5.
Figure 5a shows a diagram of the system. 
The thickness of the central layer and the pore size inside this central layer are varied, and the morphology of the central layer is optimized (the morphology of the outer layers is that of the optimized values from Fig. 4).
In other words, the values of $\alpha$, and $\tau$ from the explicit relation derived from Sevostianov and Shrestha \cite{sevostianov2010}, that maximizes the thermal efficiency is found in systems with varying middle layer thickness and middle layer pore size.

Figures 5b and 5c represent systems with a porosity of 0.5 and 0.7, respectively. The tortuosity is plotted as a function of middle layer thickness and middle layer pore radius.
As the pore radius increases (which facilitates mass transport through the porous structure) the optimized systems possess increasing tortuosity.
Similarly, in systems with higher porosity a higher tortuosity in the middle layer is predicted to maximize thermal efficiency. 
The systems favor a decrease in thermal conductivity over an increase in water flux (at least in terms of thermal efficiencies).  

Figure 6 shows the thermal efficiency for these systems. 
Figs. 6a and 6c show the optimized thermal efficiency for systems with porosities of 0.5 and 0.7, respectively, while Figs. 6b and 6d show the thermal efficiency when $\alpha = 1$ for systems with porosities of 0.5 and 0.7, respectively.
Increasing the thickness of the middle layer and the pore radius in the middle layer increases the thermal efficiency of the system, but rather than the pores in the center layer being aligned through the membrane, the optimized morphology is one in which the pores are aligned more along the membrane.
Compare the difference in thermal efficiency between Fig. 6a and 6b.
Fig. 6a is for a system whose morphology is optimized in terms of the tortuosity.
Fig. 6b is for a system with a tortuosity of 1 (straight channels through the membrane).
The difference in thermal efficiency is only around 1\% or 2\%; despite the optimized morphology possessing high tortuosity and values of $\alpha$ close to zero. 
These are relatively small variations, however, in comparison to the effects of feed temperature and porosity.
That said, while the morphology of the middle layer is predicted to play a small role in the thermal efficiency of these systems, the sheer presence of the middle layer can play a large role. In particular, Fig 6a shows that increasing the thickness of the middle layer to 80 $\mu$m (with two 10 $\mu$m layers either side) and increasing the pore radius from 0.2 $\mu$m to 2 $\mu$m (which is still captured by the combined Knudsen- and molecular-diffusion mechanisms of gas transport) can increase the thermal efficiency by around 6\%.
In Fig. 6c, showing the thermal efficiencies of system with a higher porosity of 0.7, the thermal efficiency increases by around 10\%.

\subsection{Lattice Network Morphologies}
\label{results3}

We now turn our attention to a tailored morphology. 
Figure 7a shows a lattice structure consisting of sequential layers of fibers running in perpendicular directions.
This morphology is chosen as it has a low tortuosity and will enable gas to flow easily through the straight channels, while limiting the continuity of the solid polymer phase in the direction of the temperature gradient.
This is similar to polymer membranes where the polymer is electrospun as a random network of fibers and the lattice system considered here may be a reasonable approximation of electrospun membranes; although ordered lattice structures (as considered here) can be fabricated \cite{xie2008}. 
The tortuosity is taken to be 1 and the thermal conductivity is of the form
\begin{equation}
k_m = \frac{D^2}{(L+D)^2} k_s + \frac{L^2}{(L+D)^2} k_g + \frac{2LD}{(L+D)^2} \frac{2 k_g k_s}{k_s+k_g}
\end{equation}
where $D$ is the thickness of the fibers and $L$ is the spacing between the fibers.
The thermal conductivity as a function of porosity can be shown to be of the form
\begin{equation}
k_m = (1-\epsilon)^2 k_s + \epsilon^2 k_g + 4(\epsilon - \epsilon^2) \frac{k_s k_g}{k_s + k_g}
\end{equation}
Figure 7b shows the thermal conductivity as a function of porosity, in between the Voigt-Reuss bounds.
While the Reuss bounds have a lower thermal conductivity than the lattice structure, this is generally associated with a more tortuous structure; recall the lattice structure has a tortuosity of 1.
The thermal efficiency as a function of porosity is depicted in Fig. 7c. The lattice morphology is predicted to be significantly more efficient than the Voigt limit (which also has a tortuosity of 1, but with a higher thermal conductivity).

The combination of a lattice structure and a three-layer membrane is considered in Fig. 8.
The thermal efficiency as a function of the thickness and pore radius in the middle layer is depicted for systems with a porosity of 0.5 and 0.7.
The contour scale is the same as in Fig. 6. to better compare these results with those from systems with different morphologies.
The combination of a lattice structure with a three-layer membrane significantly improves the thermal efficiency in direct contact membrane distillation.
Comparing Fig. 8 with Fig. 6, the systems with a thin middle layer and smaller pore radii show an improvement over both the systems consisting of randomly distributed interconnected phases and systems with domains aligned perpendicular to the membrane (Voigt limit).
This improvement is entirely due to the lattice structure, but the systems with thicker middle layers and larger pore radii show even further increases in thermal efficiency.
In particular, for systems with a porosity of 0.5 (comparing Fig. 8a with Figs 6a and 6b) the lattice structure is around 8\% more efficient than both systems with random and perfectly aligned morphologies. 
For systems with a porosity of 0.7 (comparing Fig. 8a with Figs 6a and 6b) this varies from around 3\% (systems with thicker and more porous middle layers) to around 10\%. 
However, to put this into perspective, the thermal efficiency in systems with three layers and a lattice morphology can be around 20\% higher than systems with single-layer membranes and randomly distributed morphologies.

\section{Summary and Conclusions}
\label{conclusions}

The effects of membrane morphology on the thermal efficiency of direct contact membrane distillation has been investigated.
Using the cross-correlation theory of Sevostianov and Shrestha \cite{sevostianov2010} we can couple the tortuosity of the porous phase with the hybridizing parameter in the hybrid Voigt-Reuss model for the thermal conductivity; the tortuosity is predicted to increase as the hybridizing parameter decreases (approaching the Reuss bound).
The variation of thermal efficiency is significant, when tortuosity and hybridizing parameter are linked, and a decrease in hybridizing parameter is predicted to be preferred at higher porosities.
In other words, Assuming a randomly distributed interconnected morphology it is found that limiting the heat conduction in direct contact membrane distillation is more important than increasing water flux for optimizing efficiency.

Introducing a three-layer membrane where the middle layer can posses larger pore sizes can increase the overall thermal efficiency of the system.
The pore size on the outer two layers is kept at 0.2 $\mu$m whilst the pore size in the middle layer is increased up to 2 $\mu$m. 
Membrane pore wetting is expected to only depend on the pore size of the outer two layers. 
Although if wetting was to occur the larger pore size in the middle layer might be expected to exacerbate the flow of feed water through the membrane.
Furthermore, these effects may only become apparent over time as the hydrophobicity of a membrane can significantly decrease over time \cite{mcgaughey2017}. 
However, the introduction of the middle layer is found to significantly increase thermal efficiency. 
Increasing the pore size in the middle layer increases the flux of water vapor without necessarily influencing the thermal conductivity of the membrane.

To maximize the thermal efficiency it is desirable to specifically design membrane morphologies that increase mass transport while minimizing heat transfer; for example, electrospun morphologies have large porosities and polymer threads are generally oriented perpendicular to the thermal gradient \cite{lalia2013}.
The direct contact membrane distillation in a system consisting of an idealized lattice structure of sequential layers of perpendicularly oriented fibers is simulated. 
This idealized structure is similar to electrospun morphologies (and it is possible to create these ideal structures experimentally). 
The low tortuosity and relatively low thermal conductivity results in high thermal efficiencies in systems with these membranes.
Furthermore, combining the idealized lattice morphology with a three-layer structure results in a membrane that significantly improves thermal efficiency by maximizing the flux of the water vapor whilst limiting thermal conductivity.
It is hoped that tailoring membrane morphology to maximize thermal efficiencies can further increase the competitiveness of direct contact membrane distillation and increase its role in meeting our future freshwater needs.

\section*{References}
\bibliographystyle{iopart-num}
\bibliography{refs}

\newpage
\section*{Figure Captions}

Figure 1: a) Different rule of mixtures for estimating the thermal conductivity as a function of porosity. b) Estimated tortuosity as a function of hybrid Voigt-Reuss parameter, derived from Sevostianov and Shrestha \cite{sevostianov2010}. Graphical illustrations of morphology as a function of the hybridizing parameter, $\alpha$, are shown.

\vspace{10pt}

Figure 2: Contour plots of thermal efficiency as a function of tortuosity and hybridizing parameter for porosities of a) 0.5, b) 0.65, and c) 0.8. The solid black line represents the estimated tortuosity as a function of hybrid Voigt-Reuss parameter, derived from Sevostianov and Shrestha \cite{sevostianov2010}.

\vspace{10pt}

Figure 3: Contour plots of thermal efficiency as a function of tortuosity and hybridizing parameter for input feed temperatures of a) 40\textsuperscript{o}C, and b) 80\textsuperscript{o}C. The solid black line represents the estimated tortuosity as a function of hybrid Voigt-Reuss parameter, derived from Sevostianov and Shrestha \cite{sevostianov2010}. c) The thermal efficiency as a function of hybridizing parameter for different temperatures.

\vspace{10pt}

Figure 4: a) Optimized tortuosity  and hybridizing parameter, maximizing thermal efficiency, as a function of porosity. b) Optimized thermal efficiency as a function of porosity. The thermal efficiency of systems using the Voigt-Reuss bounds are included for comparison.

\vspace{10pt}

Figure 5: a) Schematic of triple-layer system. The thickness and pore radius of the middle layer are varied. The optimized tortuosity of the middle layer as a function of the thickness and pore radius of the middle layer for porosities of b) 0.5 and c) 0.7.

\vspace{10pt}

Figure 6: Thermal efficiency as a function of middle layer pore size and thickness. a) optimized thermal efficiency for system with a porosity of 0.5, b) thermal efficiency for system with $\alpha = 1$ and a porosity of 0.5, c) optimized thermal efficiency for system with a porosity of 0.7, and d) thermal efficiency for system with $\alpha = 1$ and a porosity of 0.7.

\vspace{10pt}

Figure 7: a) A schematic of the lattice structure is shown. b) The thermal conductivity of the lattice membrane as a function of porosity is contrasted with the Voigt-Reuss bounds. c) The thermal efficiency as a function of porosity in systems with a lattice morphology  are contrasted with systems with a Voigt morphology (both with a tortuosity of 1).

\vspace{10pt}

Figure 8: The thermal efficiency of a three-layer membrane with lattice structure is depicted as a function of the thickness of the middle-layer and the pore radius of the middle layer. 
Systems with a porosity of a) 0.5 and b) 0.7 are shown.
In contrast with Fig. 6, the lattice structure in these systems results in improved thermal efficiency, in addition to the increases that can be seen for thicker middle layers with larger pore sizes.

\newpage
 \pagenumbering{gobble} 

\begin{figure}
	\begin{center}
		\includegraphics[width=0.8\linewidth]{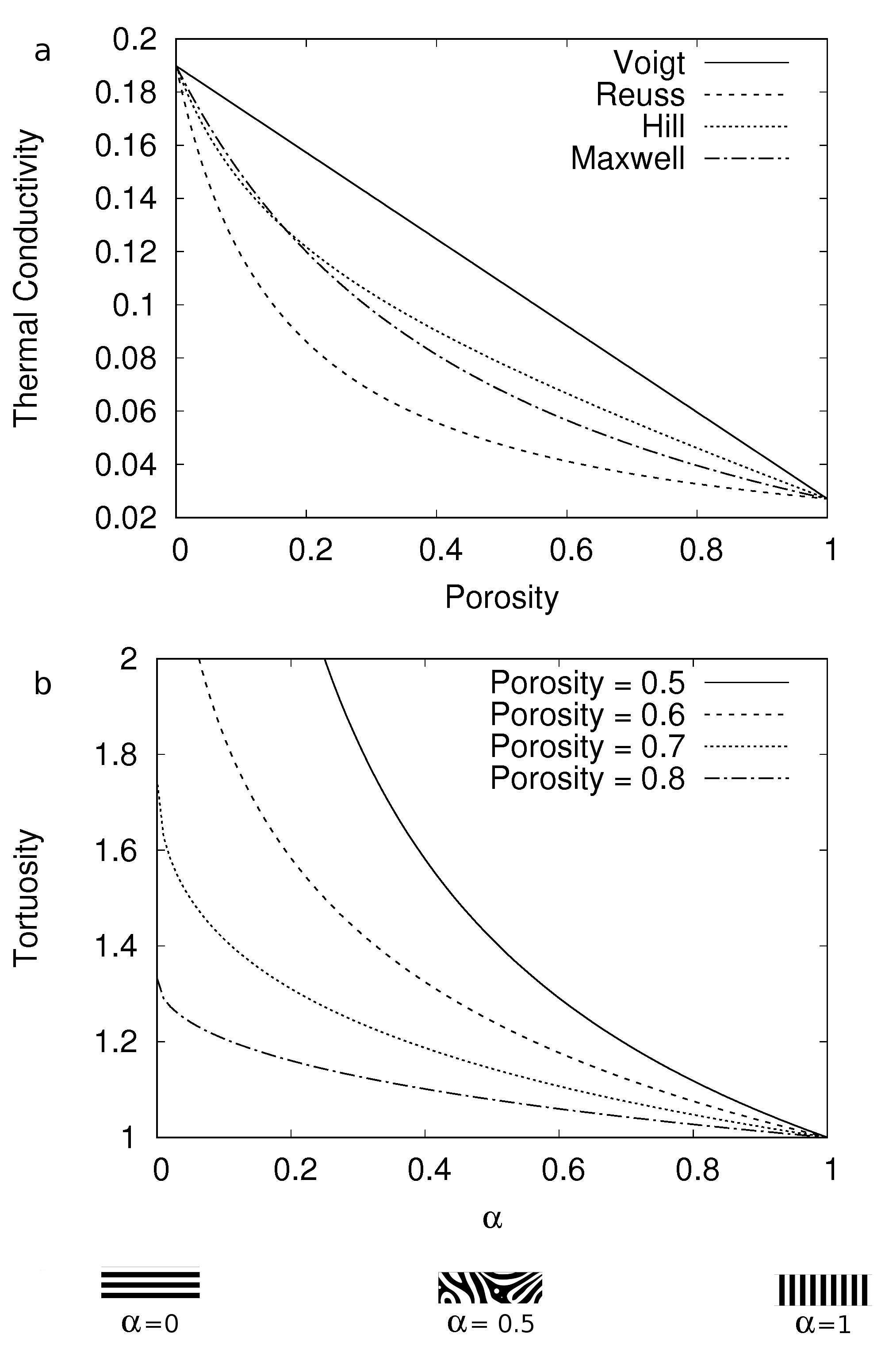}
	\end{center}
	  \caption{}
\end{figure}

\newpage

\begin{figure}
	\begin{center}
		\includegraphics[width=0.65\linewidth]{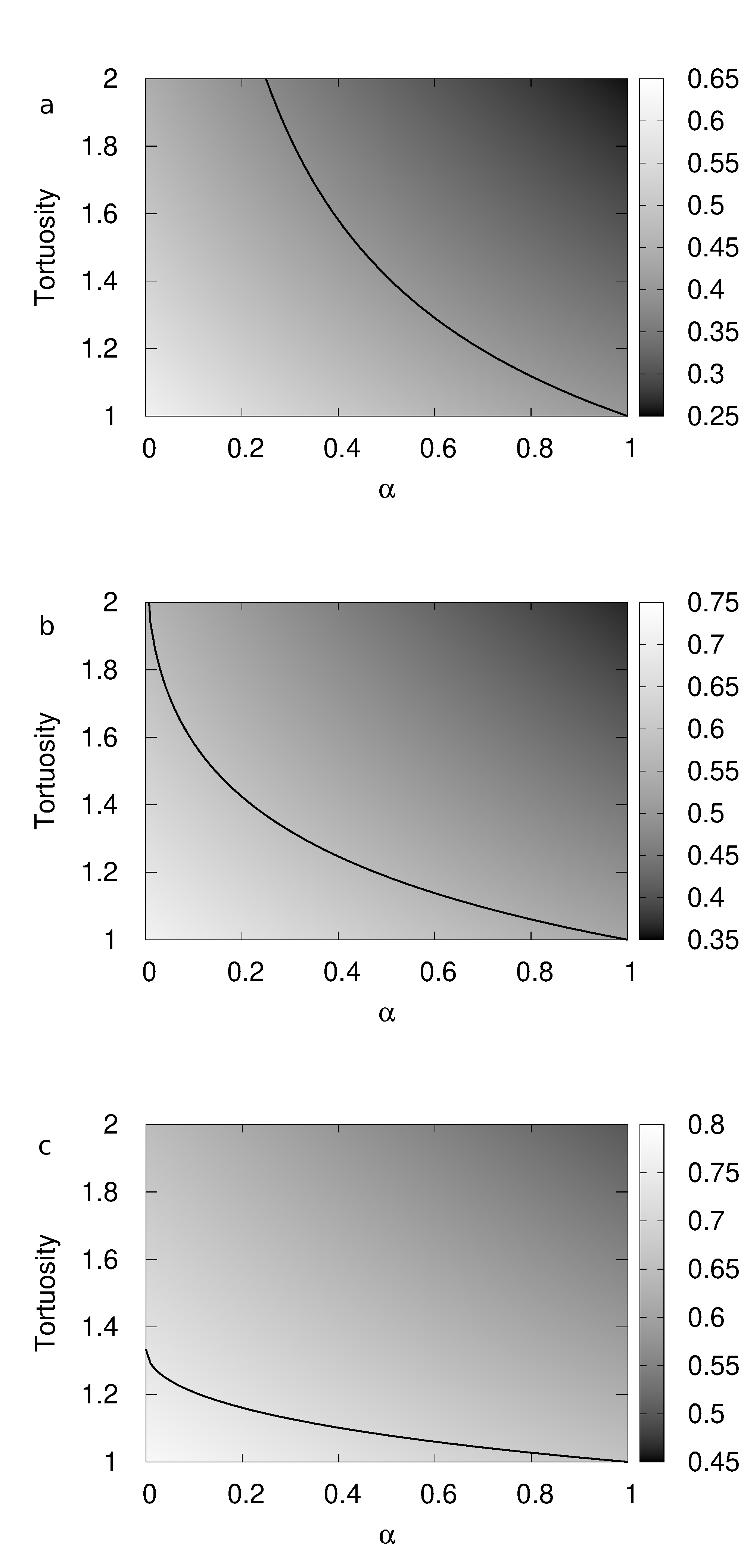}
	\end{center}
	  \caption{}
\end{figure}

\newpage

\begin{figure}
	\begin{center}
		\includegraphics[width=0.65\linewidth]{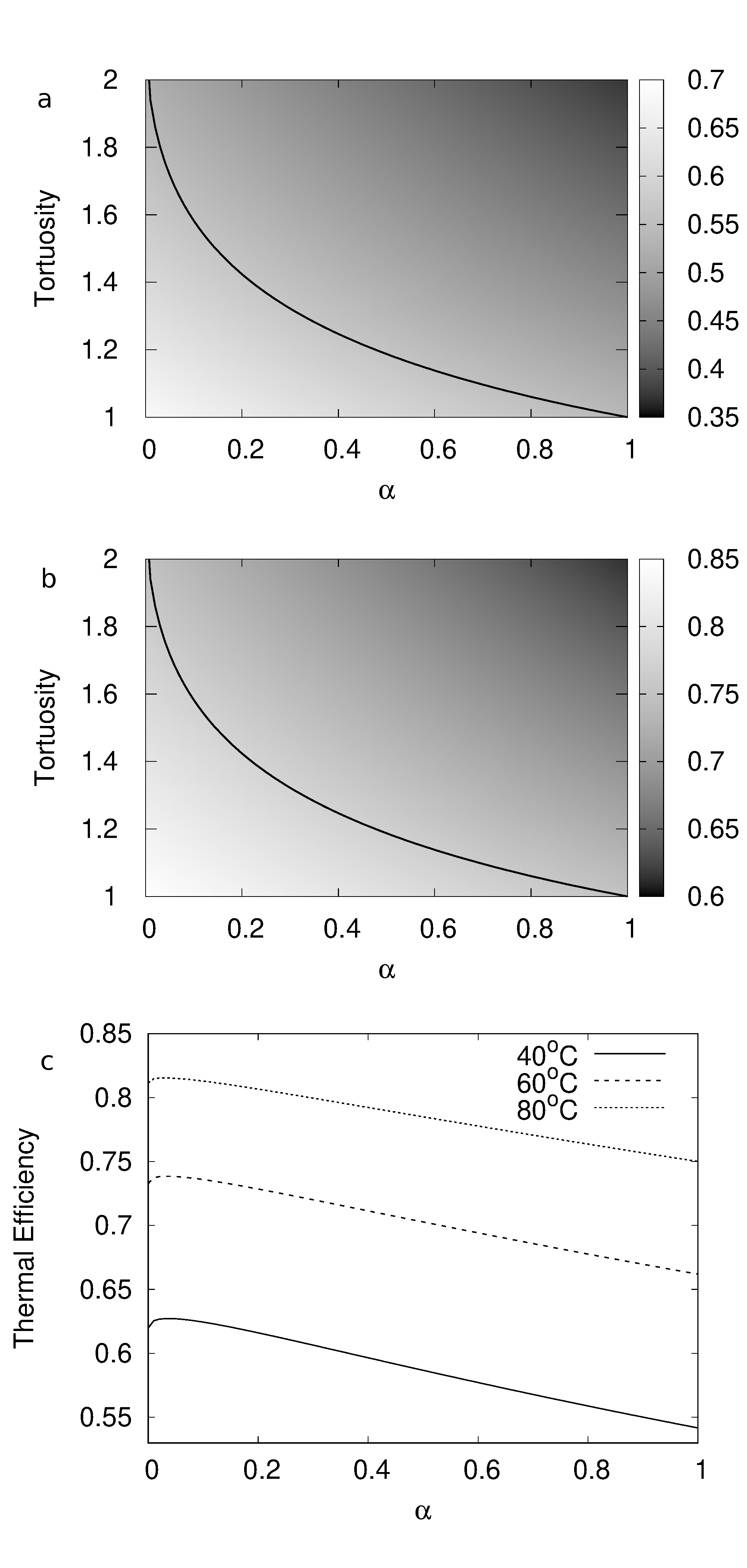}
	\end{center}
	  \caption{}
\end{figure}

\newpage

\begin{figure}
	\begin{center}
		\includegraphics[width=0.75\linewidth]{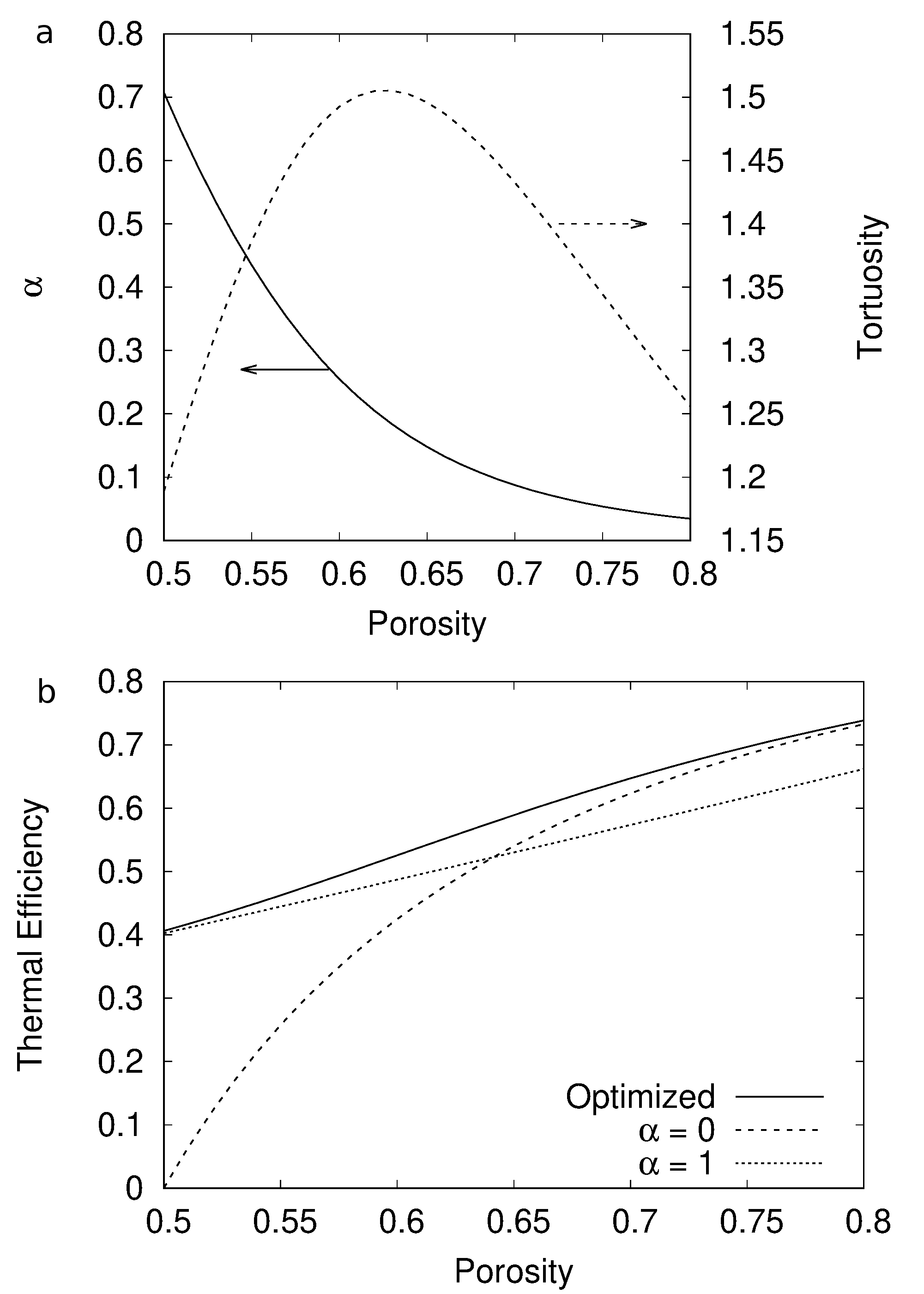}
	\end{center}
	  \caption{}
\end{figure}

\newpage

\begin{figure}
	\begin{center}
		\includegraphics[width=0.75\linewidth]{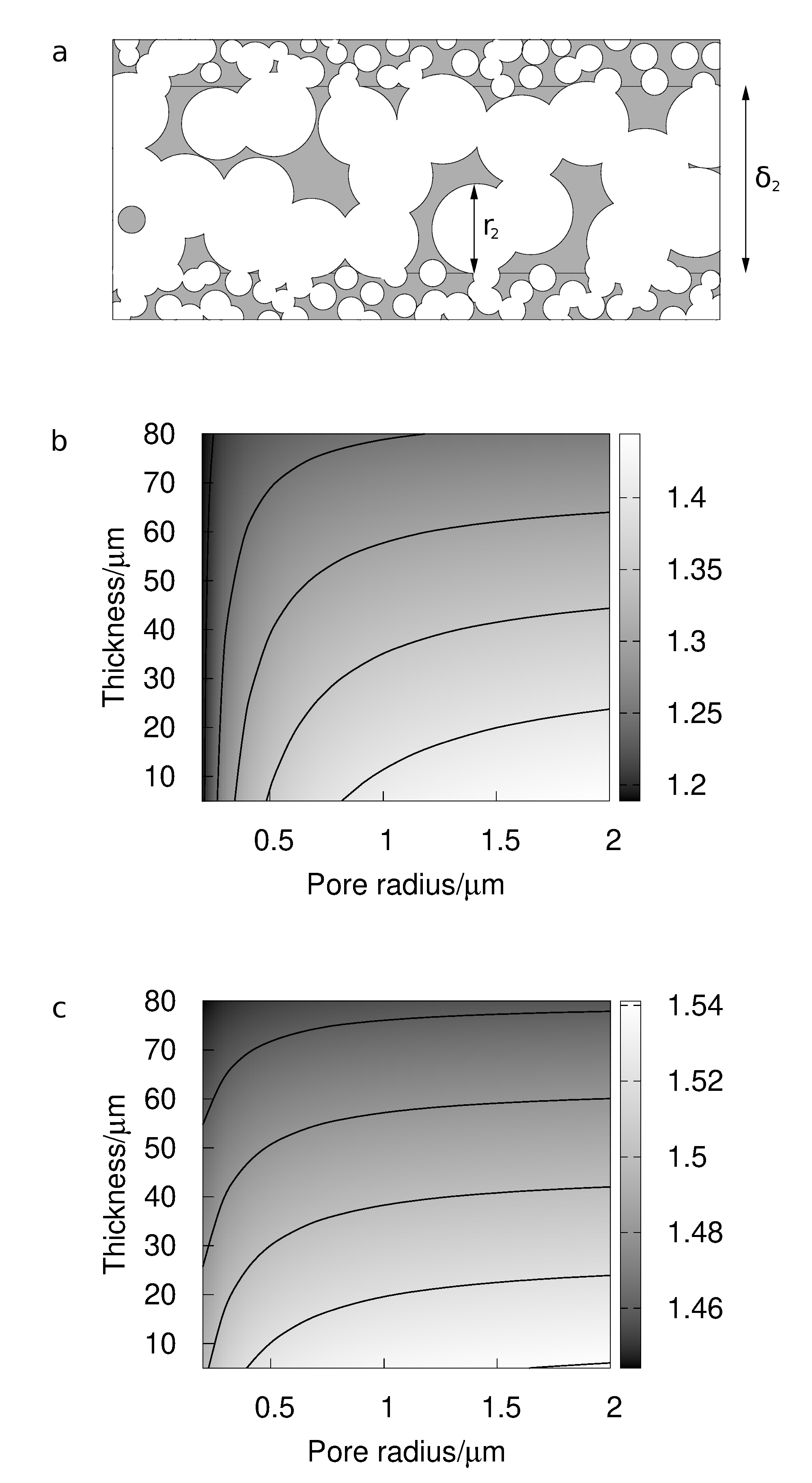}
	\end{center}
	  \caption{}
\end{figure}

\newpage

\begin{figure}
	\begin{center} 
		\makebox[\textwidth][c]{\includegraphics[width=1.3\linewidth]{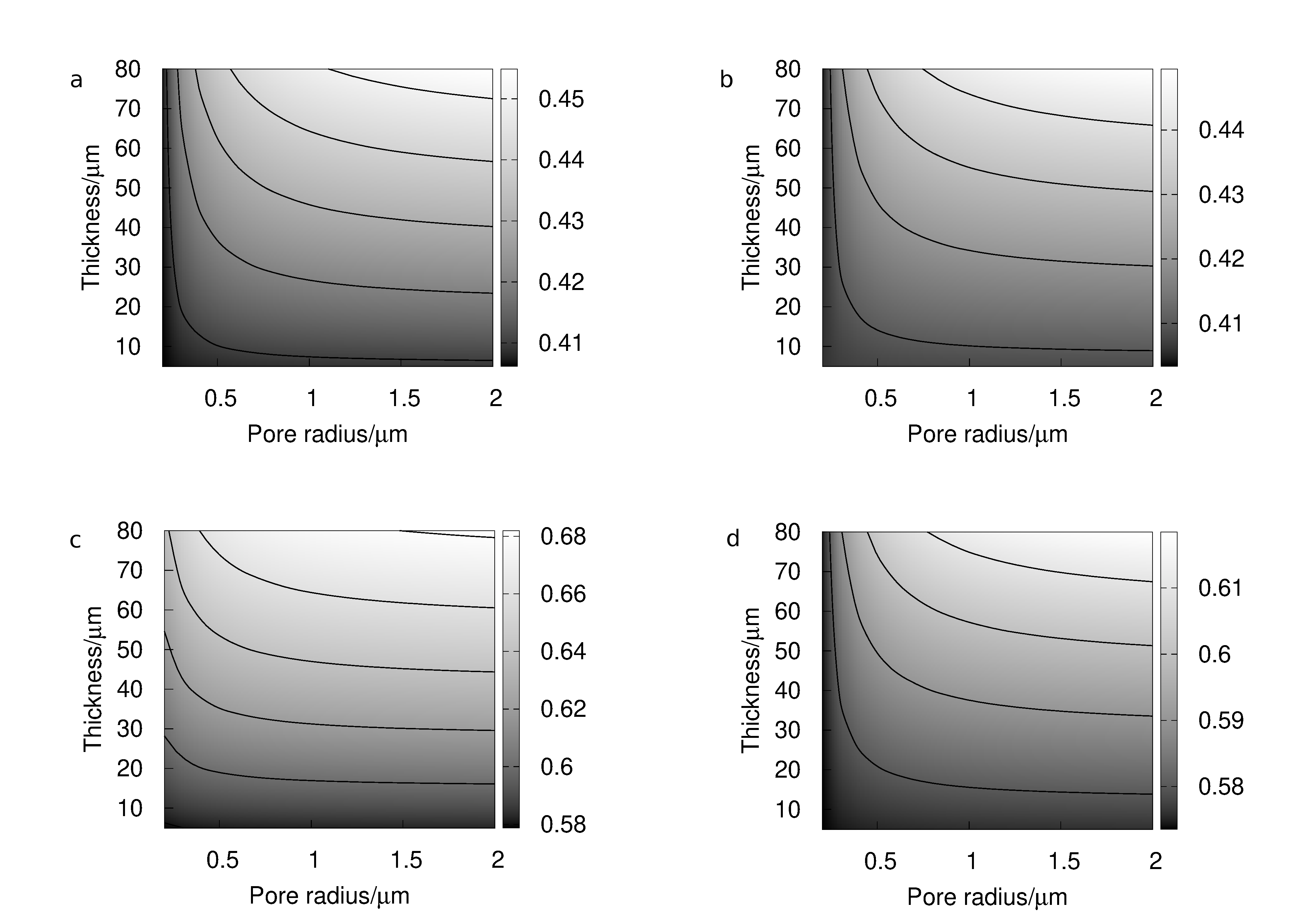}}%
		
	\end{center}
	  \caption{}
\end{figure}

\newpage

\begin{figure}
	\begin{center}
		\includegraphics[width=0.7\linewidth]{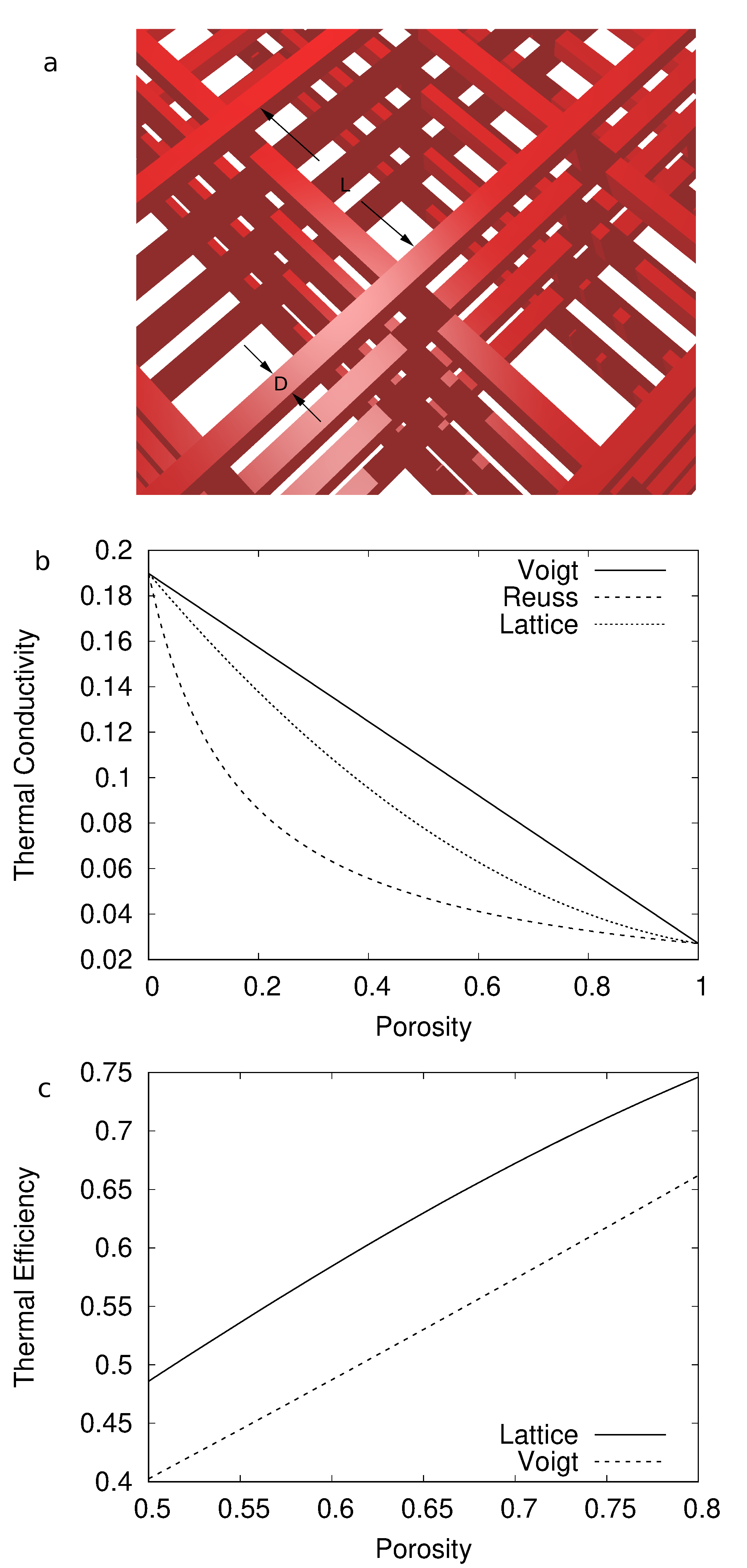}
	\end{center}
	  \caption{}
\end{figure}

\newpage

\begin{figure}
	\begin{center}
		\includegraphics[width=0.8\linewidth]{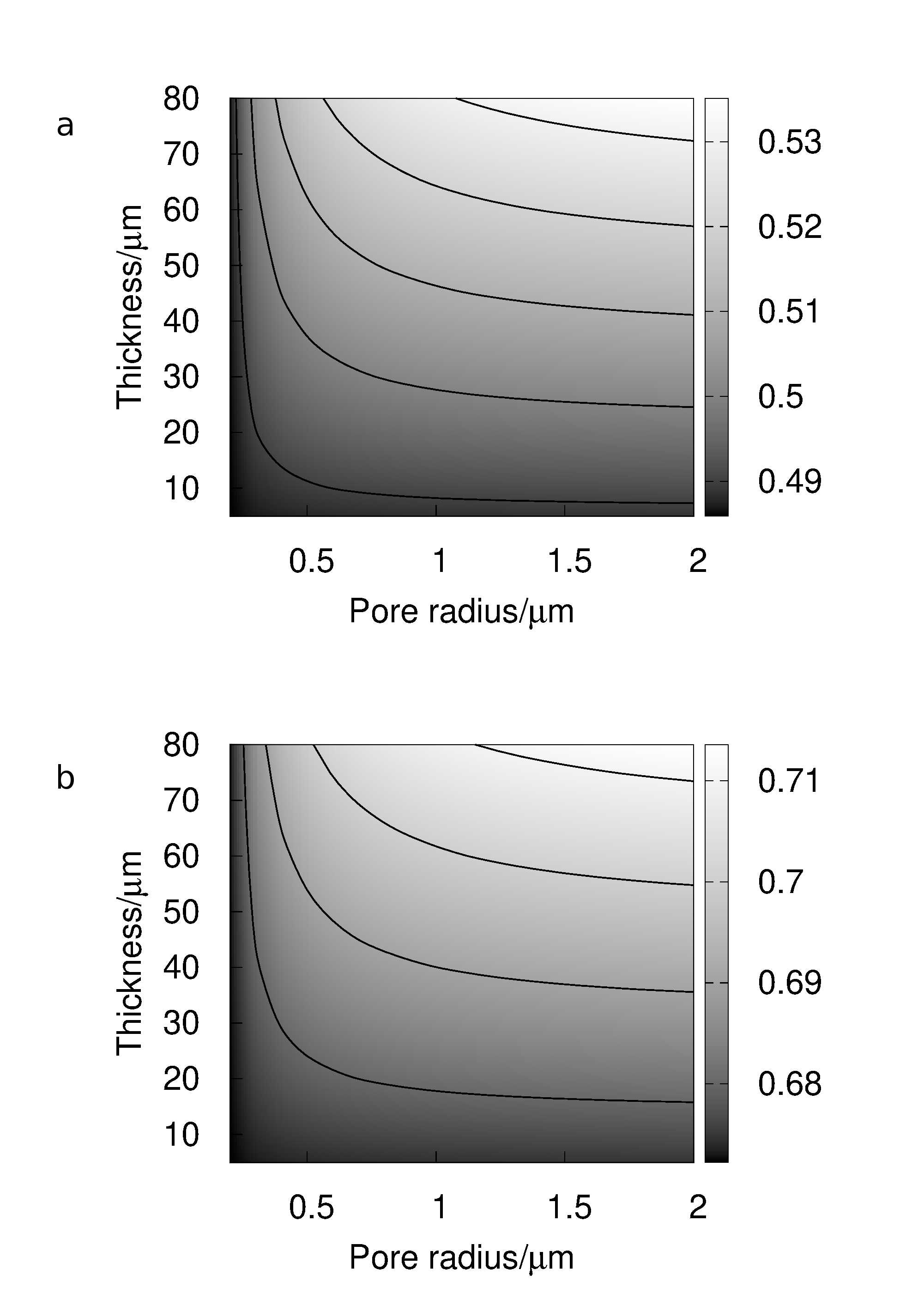}
	\end{center}
	  \caption{}
\end{figure}

\end{document}